Reconsidering the asymptotic null distribution of likelihood ratio tests for genetic linkage in multivariate variance components models

Summer S. Han[#]
Joseph T. Chang

[#]Corresponding Author

Summer S. Han
Department of Statistics
Yale University
24 Hillhouse Avenue
New Haven, CT 06520
Phone: (203)432-0666
Fax: (203)432-0633
seongmin.han@yale.edu

Joseph T. Chang
Department of Statistics
Yale University
24 Hillhouse Avenue
New Haven, CT 06520
Phone: (203)432-0642
Fax: (203)432-0633
joseph.chang@yale.edu

## Abstract

Accurate knowledge of the null distribution of hypothesis tests is important for valid application of the tests. In previous papers and software, the asymptotic null distribution of likelihood ratio tests for detecting genetic linkage in multivariate variance components models has been stated to be a mixture of chi-square distributions with binomial mixing probabilities. Here we show, by simulation and by theoretical arguments based on the geometry of the parameter space, that all aspects of the previously stated asymptotic null distribution are incorrect—both the binomial mixing probabilities and the chi-square components. Correcting the null distribution gives more conservative critical values than previously stated, yielding P values that can easily be ten times larger. The true mixing probabilities give the highest probability to the case where all variance parameters are estimated positive, and the mixing components show severe departures from chi-square distributions. Thus, the asymptotic null distribution has complex features that raise challenges for the assessment of significance of multivariate linkage findings. We propose a method to generate an asymptotic null distribution that is much faster than other empirical methods such as gene-dropping, enabling us to obtain P values with higher precision more efficiently.

# 1.    Introduction and main results

Variance components methods are widely used for mapping quantitative trait loci (Almasy and Blangero, 1998; Amos, 1994; Goldgar, 1990; Schork, 1993; SOLAR, 2008).  Multivariate extensions of variance component methods, which simultaneously test genetic effects on multiple traits by exploiting extra information related to correlations among the traits, have attracted much attention and have been found to be more powerful than univariate methods by several studies (Almasy *and others*, 1997; Amos *and others*, 2001; Boomsma and Dolan, 1998; Schmitz *and others*, 1998; Williams *and others*, 1999a; Williams *and others*, 1999b).

The asymptotic null distribution of a univariate variance components test, which is a likelihood ratio test for comparing the full model to the null model where the genetic effect variance parameter is constrained to zero, is well known to be a 50:50 mixture of a point mass at zero and $\chi_1^2$ (the chi-squared distribution with 1 degree of freedom).  This nonstandard limiting distribution arises because the parameter of interest lies on the boundary of the parameter space under the null hypothesis (Self and Liang, 1987).

An analogous mixture distribution was applied to multivariate variance components tests by a number of papers (Almasy *and others*, 1997; Amos *and others*, 2001; Amos and de Andrade, 2001; Kraft *and others*, 2004; Williams *and others*, 1999a; Williams *and others*, 1999b) in which the asymptotic null distribution of the tests was stated to be a mixture of a point mass at zero and several chi-squared distributions, with binomial mixing probabilities.  This multivariate method has become widely available; for example, it is implemented in SOLAR (Almasy and Blangero, 1998; SOLAR, 2008), one of the most commonly used software packages for mapping quantitative loci.  However, the evaluation and verification of the asymptotic null distribution of this test have not yet been done systematically.   As part of a study comparing the power of univariate to bivariate variance components methods, Amos *and others* (2001) conducted a simulation of the null distribution. Their bivariate simulation result was not consistent with the previously stated asymptotic null distribution that they assumed, but the source and nature of the inconsistency have remained unexplained (Amos and de Andrade, 2001).

Here we study the null distribution of multivariate variance components tests.  We demonstrate that all aspects of the previously stated asymptotic null distribution are incorrect—both the binomial mixing probabilities and the chi-squared components. We show this both by simulation and by theoretical arguments based on the geometry of the parameter space.  The true mixing probabilities give the highest probability to the case where all variance parameters are estimated positive, and each mixing component shows severe departure from chi-square distributions.

Thus, the asymptotic null distribution of the test under consideration has complicated features that hinder assessment of significance in multivariate linkage findings.  We propose a method to assess significance by generating the asymptotic null distribution making use of the Fisher information estimated from given data. In an example application to an illustrative data set, we compare our method to other well known empirical methods such as gene-dropping, permutation and bootstrap methods, and show that our method is much faster and hence useful for assessing P values with higher precision

## 2. Background

### 2.1. *Variance component linkage methods*

Variance components linkage methods partition the variance of the traits into several random effect components, one of which captures the influence of a hypothesized "major gene" that affects the traits. Using a model that assumes multivariate normal distributions, a likelihood ratio test for a major gene effect located at a certain map position is performed by comparing the likelihood under the null model where the major gene effect variance parameter is constrained to be zero to the likelihood under the full model where the parameter is estimated without restriction.

In univariate tests, the response variable includes one measured trait for each individual, modeled as a sum of three independent normally distributed random effects: an additive major gene effect, which is to be tested on each genetic marker, a polygenic effect, and an environmental effect with corresponding variance parameters $a^2$, $g^2$ and $e^2$ respectively. For simplicity, here let us assume one sib-pair in each family. The trait vector for the $i$ th family follows a multivariate normal distribution:

$$y_i = (y_{i1,} \ y_{i2}) \sim N(\mu, \Sigma_i) \ \text{ with } \ \Sigma_i = \begin{bmatrix} a^2 + g^2 + e^2 & \pi_{i,12}a^2 + 2\phi_{i,12}g^2 \\ \pi_{i,12}a^2 + 2\phi_{i,12}g^2 & a^2 + g^2 + e^2 \end{bmatrix}.$$

Here the identity by descent (IBD) sharing proportion $\pi_{i,12}$ quantifies the allele sharing between individuals 1 and 2 at the locus being tested; $\pi_{i,12}$ is pre-estimated using marker data and treated as fixed during maximization of the likelihood function. The kinship coefficient $\phi_{i,12}$ between individuals 1 and 2 is obtained from the pedigree without using any genetic marker information; for example, for a sib-pair, $\phi_{i,12} = \frac{1}{4}$. Linkage is tested by comparing the null hypothesis $H_0 : a^2 = 0$ to the alternative $H_1 : a^2 > 0$. The likelihood-ratio test statistic (LRT hereafter) is twice the difference between the log-likelihood of the full model and the model restricted according to the null hypothesis. The parameter $a^2$, which must be nonnegative, lies on the boundary of the parameter space under the null hypothesis $a^2 = 0$, so that a nonstandard boundary condition applies (Self and Liang, 1987), and the asymptotic null distribution of LRT is $\frac{1}{2}\chi_0^2 + \frac{1}{2}\chi_1^2$, that is, the mixture distribution of 50% point mass at zero (equivalent to $\chi_0^2$) and 50 % $\chi_1^2$.

Multivariate variance components models are a natural extension of the above single-trait model, where the response variables include $k$ measured traits for each individual. Continuing to illustrate with the case of one sib-pair per family, $y_i = (y_{i11}, y_{i12}, \ldots, y_{i1k}; y_{i21}, y_{i22}, \ldots, y_{i2k}) \sim N(\mu, \Sigma_i)$, where $y_{ijt}$ is the value of trait $t$ measured on individual $j$ in family $i$, and

$$\Sigma_i = \begin{bmatrix} A + G + E & \pi_{i,12}A + 2\phi_{i,12}G \\ \pi_{i,12}A + 2\phi_{i,12}G & A + G + E \end{bmatrix}. \tag{1}$$

Here $G = \left(g_{ij}\right)$ and $E = \left(e_{ij}\right)$ are general $k \times k$ covariance matrices for the polygenic effect and environmental effect, respectively. The covariance matrix for the additive major gene effect, $A = \left(a_{ij}\right)$, is assumed to be an outer product of the form $A = \begin{pmatrix} a_1 & a_2 & \cdots & a_k \end{pmatrix}^T \begin{pmatrix} a_1 & a_2 & \cdots & a_k \end{pmatrix}$. Here the variances, which are the diagonal entries $a_1^2, \ldots, a_k^2$, are of course nonnegative, but the entries $a_i$ can take both positive and negative values, so that the covariances $a_{ij} = a_i a_j$ can be positive or negative. The null hypothesis to be tested, $H_0 : a_1^2 = \cdots = a_k^2 = 0$, is that all of the major gene effect variances are 0.

This model, also known as the *single factor model* (Evans *and others*, 2004; Vogler *and others*, 1997) or the *complete pleiotropic model* (Almasy *and others*, 1997; Amos *and others*, 2001; Amos and de Andrade, 2001; Kraft *and others*, 2004; Williams *and others*, 1999a; Williams *and others*, 1999b), arises when the dominance components of variance for the effects of a single major gene on each trait are assumed to be 0. The idea of the assumed form for $A$ is that the effect on a given trait of a given major gene genotype is a proportionality constant that depends on the trait times an effect size that depends on the genotype. That is, the major gene genotype is modeled as a latent factor that affects the traits in a linear pleiotropic manner. The single factor model has been extensively used (Amos *and others*, 2001; Amos and de Andrade, 2001; Evans, 2002; Evans and Duffy, 2004; Evans *and others*, 2004; Marlow *and others*, 2003; Monaco, 2007; Vogler *and others*, 1997). For example, applications of this model to analysis of multivariate linkage data in particular disorders include Marlow *and others* (2003) for dyslexia and Monaco (2007) for specific language impairment. The model has also been used for studying the power of bivariate linkage by, for example, Amos *and others*(2001), Evans (2002), and Evans and Duffy (2004).

## 2.2. Asymptotic null distributions and binomial mixing probabilities

The null hypothesis of the likelihood ratio test in the models discussed above violates the regularity conditions that imply the typical asymptotic chi-square distribution, the most obvious violation being that constraining a variance to be 0 forces the variance to lie in the boundary of the parameter space rather than its interior. In previous work *and others*, 2001; SOLAR, 2008), the asymptotic null distribution of the multivariate LRT statistic for testing $k$ traits has been stated to be a mixture distribution of point mass at zero and chi-square distributions with degrees of freedom from 1 to $k$, with mixing probabilities coming from the Binomial distribution $B\left(k, \dfrac{1}{2}\right)$, that is,

$$\frac{1}{2^k}\binom{k}{0}\chi_0^2 + \frac{1}{2^k}\binom{k}{1}\chi_1^2 + \ldots + \frac{1}{2^k}\binom{k}{k-1}\chi_{k-1}^2 + \frac{1}{2^k}\binom{k}{k}\chi_k^2. \tag{2}$$

For $k =2$, for example, this asymptotic null distribution would be $0.25\chi_0^2 + 0.5\chi_1^2 + 0.25\chi_2^2$. The reasoning in that previous work (Amos *and others*, 2001) was that, under the truth of the null hypothesis $H_0 : a_1^2 = a_2^2 = 0$, both parameters $a_1^2$ and $a_2^2$ are estimated to be positive in ¼ of the parameter space, one out of the two parameters $a_1^2$ and $a_2^2$ is estimated to be positive (and the other is estimated to be 0) in ½ of the parameter space, and neither of the two parameters $a_1^2$ and $a_2^2$ is estimated to be positive in ¼ of the parameter space. Table 1 shows mixing probabilities and critical values according to formula (2) for significance levels $\alpha =0.01$ and 0.05 for $k$ - trait asymptotic null distributions for $k = 2, 3, 4,$ and 5.

Table 1.  Mixing probabilities and critical values of the previously stated asymptotic null distributions for $k$-trait multivariate tests according to the binomial mixture formula (2). Critical values $C_k^*$ were calculated so that the weighted tail probabilities of $k$ chi-square distributions would sum to $\alpha$ for the given quantile $C_k^*$.  For example for $k = 2$, $C_2^*$ was chosen to satisfy $\alpha = 0.5*(1-\text{pchisq}(C_2^*,\text{df=1})) + 0.25*(1-\text{pchisq}(C_2^*,\text{df=2}))$, where pchisq represents the chi-square cumulative distribution function.

| | $\chi_0^2$ | $\chi_1^2$ | $\chi_2^2$ | $\chi_3^2$ | $\chi_4^2$ | $\chi_5^2$ | Critical Value ($\alpha$=0.01) | Critical Value ($\alpha$=0.05) |
|---|---|---|---|---|---|---|---|---|
| $k$=2 | 0.2500 | 0.5000 | 0.2500 | | | | 7.289 | 4.231 |
| $k$=3 | 0.1250 | 0.3750 | 0.3750 | 0.1250 | | | 8.746 | 5.435 |
| $k$=4 | 0.0625 | 0.2500 | 0.3750 | 0.2500 | 0.0625 | | 10.019 | 6.498 |
| $k$=5 | 0.0313 | 0.1563 | 0.3125 | 0.3125 | 0.1563 | 0.0313 | 11.183 | 7.480 |

## 3.    Simulations

### 3.1.    Methods

We simulated the null distributions of $k$-trait multivariate LRT statistics for $k = 2,3,4,5$.  For each test, we generated 1000 datasets each including 2000 independent sib-pairs with traits simulated under the truth of the null hypothesis $H_0 : a_1^2 = \cdots = a_k^2 = 0$ using the multivariate normal distributions specified at (1). Each trait had total variance 1 with $g_{11} = \cdots = g_{kk} = 0.4$ (i.e. the polygenic effect explained 40% of the total variance of each trait) and $e_{11} = \cdots = e_{kk} = 0.6$ (i.e. the rest of the variance was due to environmental effects). Polygenic and environmental correlation coefficients among traits were assigned the value 0.1, so that $g_{ij} = 0.04$ and $e_{ij} = 0.06$ for $i \neq j$. The IBD sharing levels $\pi_{i,12}$ were simulated from s with probabilities {¼, ½, ¼} respectively, which amounts to assuming complete linkage information and random sampling.

Optimization was done using the Mx software (Neale *and others*, 1999). We recorded the LRT statistic and estimates $\hat{a}_1^2,...,\hat{a}_k^2$ for each replicate dataset. In order to separate the mixture distribution of LRT into its components, we grouped the replicate results by the number of major gene variance parameters that were estimated positive, which we will denote by $\nu$. There were $k+1$ groups, starting from zero (where none of the $k$ parameters was estimated positive) up to $k$ (where all of the $k$ parameters were estimated positive), so that $\nu$ took its values in $\{0,1,2,\ldots,k\}$.

Correctly deciding the number of variance parameters that were estimated positive is important in this paper.  Simply counting the estimates that were recorded exactly as zero by Mx would give incorrect counts, since the computations done by Mx are of limited precision.  Many estimates were reported by Mx as not exactly zero but quite close to zero, e.g. $1 \times 10^{-9}$, and a procedure was required to classify them.  To determine whether a given variance parameter should be estimated as zero or not, we performed an optimization over a partially restricted model in which only the given variance parameter is restricted to zero.  If the optimized log-likelihood value of this partially constrained model was strictly less than the optimized log-likelihood for the full model, we categorized the estimate of the given variance parameter as non-zero (i.e. positive); if the log-likelihood value of the full model was the same as that of the partially constrained model, then we counted the variance estimate as zero. This way, for each replicate, we ran an additional $k$ partially restricted model optimizations to decide individually whether each variance parameter was zero.

After grouping the replicate results by the number of parameters that were estimated positive, we obtained the empirical distribution of LRT statistics in each mixing component group, and the empirical mixing probabilities were calculated simply as the fraction of each group within the total number of replicates. In order to check whether the distribution within each component followed a $\chi^2$ distribution with specified degrees of freedom, we conducted Kolmogorov-Smirnov tests.

### 3.2. Results

Table 2, which shows the results of the null distribution simulations, reveals that the mixing probabilities do not agree with the probabilities anticipated in Table 1, that is, the binomial distribution previously used in the literature and software. For example for $k = 2$, the probabilities for $\nu$ (the number of variance parameters being estimated positive) taking the values 0, 1, and 2 are 0.142, 0, and 0.858, respectively, whereas the corresponding binomial probabilities are 0.25, 0.5, and 0.25. The substantial discrepancies already seen in the case $k = 2$ become larger for higher $k$. Instead of a binomial distribution having probabilities $\binom{k}{\nu} 2^{-k}$ for $\nu = 0, 1, 2, \ldots, k$, the true mixing distribution puts positive probability only on $\nu = 0$ and $\nu = k$. Furthermore, rather than having equal probabilities for the cases $\nu = 0$ and $\nu = k$, the true distribution puts larger probability on $\nu = k$, and this effect becomes more extreme as $k$ increases.

Table 2. Mixing probabilities, Kolmogorov-Smirnov tests, and critical values for $k$ - trait multivariate test null distributions obtained by simulations. The first six columns are probabilities for various values for $\nu$, calculated as the fraction of replicates that yielded that value for $\nu$. The Kolmogorov tests were conducted for the $\nu = k$ cases.

| | Number of major gene effect variance parameters estimated positive ($\nu$) | | | | | | Kolmogorov-Smirnov test (P-value) | Critical value | |
|---|---|---|---|---|---|---|---|---|---|
| | 0 | 1 | 2 | 3 | 4 | 5 | | $\alpha$ =0.01 | $\alpha$ =0.05 |
| $k$ =2 | 0.142 | - | 0.858 | | | | $3.3 \times 10^{-12}$ | 8.640 | 5.485 |
| $k$ =3 | 0.020 | - | - | 0.980 | | | $4.4 \times 10^{-9}$ | 12.221 | 7.696 |
| $k$ =4 | 0.001 | - | - | - | 0.999 | | $8.2 \times 10^{-4}$ | 14.820 | 10.172 |
| $k$ =5 | - | - | - | - | - | 1.000 | $2.4 \times 10^{-14}$ | 15.693 | 12.458 |

Figure 1 shows empirical distributions of the LRT statistics in the cases where $\nu = k$. We focused on these cases since, as shown in Table 2, only the $\nu = k$ component had large enough counts to be drawn as histogram. The blue curves show the $\chi_k^2$ density functions; according to the previously stated asymptotic null distribution, one would anticipate the empirical distributions would be close to $\chi_k^2$. Figure 1 reveals some departures from the anticipated $\chi^2$ distributions. In $k = 2$ and 3, there are more points near zero than anticipated, in $k$ =5, the distribution is shifted to the right compared to the $\chi_5^2$ curve, and $k$ =4 is somewhat in between these two different patterns. To quantify the departure from the anticipated $\chi_k^2$ distributions, we conducted Kolmogorov-Smirnov tests, and the results are shown in Table 2. From the small P-values we conclude that the empirical distributions do not follow the anticipated $\chi_k^2$ distributions.

Figure 1. Empirical null distributions of LRT statistics for $k$-trait multivariate tests for the cases where $\nu = k$. The blue curve is the $\chi_k^2$ density.

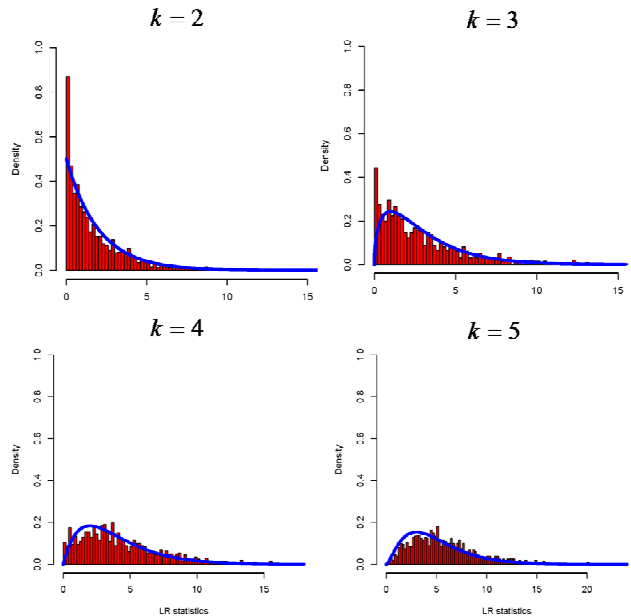

The last two columns in Table 2 are critical values for the significance levels $\alpha = 0.01$ and $0.05$. Compared to the critical values anticipated in Table 1, these critical values are larger, that is, more conservative. This implies the previously stated null distribution can lead to false positive findings. For example, suppose we observe a LRT statistic value 14 in a multivariate test with $k = 5$ traits, which the previously stated null distribution assigns a P-value of 0.002849. According to the simulated null distribution, however, the true P-value is 0.025, which is about ten times less significant.

## 4. Geometric explanations and applications

Our simulation results contain a number of interesting features: (i) zero mixing proportion for the case where only a subset of boundary parameters are estimated positive, and highest proportion on the case where all variance parameters are estimated positive, (ii) departure from chi-square distributions within the mixing components. Here we explain these results using theoretical arguments based on the geometry of the parameter space.

The likelihood ratio test of the null hypothesis $\theta \in \Omega_0$ versus the alternative hypothesis $\theta \in \Omega_1$ uses the LRT statistic $2\log\left(\sup_{\theta \in \Omega_1} L(\theta)\right) - 2\log\left(\sup_{\theta \in \Omega_0} L(\theta)\right)$, where $L$ is the likelihood function. According to Theorem 3 in Self and Liang (Self and Liang, 1987), when $\theta = \theta_0$, under certain regularity conditions, the asymptotic distribution of LRT is the same as the distribution of the LRT for testing $\theta \in C_{\Omega_0}$ versus $\theta \in C_{\Omega_1}$ based on a single observation $Y$, where $Y \sim N\left(\theta, \, I^{-1}(\theta_0)\right)$ and $C_{\Omega_0}$ and $C_{\Omega_1}$ are cones approximating $\Omega_0$ and $\Omega_1$ at $\theta_0$. This latter LRT takes the form

$$\inf_{\theta \in C_{\Omega_0}} (Y - \theta)^T I(\theta_0)(Y - \theta) - \inf_{\theta \in C_{\Omega_1}} (Y - \theta)^T I(\theta_0)(Y - \theta),$$

so that, defining $Z = Y - \theta_0$, the LRT is equivalent

$$LRT = \inf_{\theta \in C_{\Omega_0} - \theta_0} (Z - \theta)^T I(\theta_0)(Z - \theta) - \inf_{\theta \in C_{\Omega_1} - \theta_0} (Z - \theta)^T I(\theta_0)(Z - \theta), \tag{3}$$

where $C_{\Omega_0} - \theta_0 = \left\{ \theta - \theta_0 : \theta \in C_{\Omega_0} \right\}$ and $Z \sim N\left(0, \, I^{-1}(\theta_0)\right)$ under the null hypothesis. For the special case when $I(\theta_0)$ is the identity matrix $I$,

$$LRT = \inf_{\theta \in C_{\Omega_0} - \theta_0} \|Z - \theta\|^2 - \inf_{\theta \in C_{\Omega_1} - \theta_0} \|Z - \theta\|^2. \tag{4}$$

So in the case $I(\theta_0) = I$, LRT is the difference between the squared (Euclidean) distance from $Z$ to $C_{\Omega_0} - \theta_0$ and squared distance from $Z$ to $C_{\Omega_1} - \theta_0$.

Below, we will mainly write expressions explicitly for the case of $k = 2$ traits with the information matrix $I(\theta_0)$ being the identity matrix, which reveals the essential ideas with minimal notational complexity.

### 4.1.  Mixing probabilities

A way of thinking about the problem that is tempting but leads to incorrect conclusions, including the asymptotic distribution incorrectly stated in previous papers, is as follows. Denote the parameters of interest by $\theta = (\theta_1, \theta_2) = (a_{11}, a_{22}) = \left(a_1^2, a_2^2\right)$. The null hypothesis $\Omega_0$ is $\theta = (0, 0)$, and the alternative $\Omega_1$ is the first quadrant $\theta \in \left\{ (\theta_1, \theta_2) : \theta_1 > 0, \, \theta_2 > 0 \right\}$, and the approximating cones $C_{\Omega_0}$ and $C_{\Omega_1}$ are the same as $\Omega_0$ and $\Omega_1$. Let $Z \sim N(0, I)$ and

$LRT = \inf_{\theta \in \Omega_0} \|Z - \theta\|^2 - \inf_{\theta \in \Omega_1} \|Z - \theta\|^2 = \|Z\|^2 - \inf_{\theta \in \Omega_1} \|Z - \theta\|^2 = \|Z\|^2 - \left\|Z - \hat{\theta}\right\|^2$, where $\hat{\theta} = \left(\hat{\theta}_1, \, \hat{\theta}_2 \, \right)$ is the vector $\theta$ that achieves the infimum in $\inf_{\theta \in \Omega_1} \|Z - \theta\|^2$, that is, the projection of $Z$ onto the first quadrant $\Omega_1$. The random point $Z = (Z_1, \, Z_2)$ falls in each of the four quadrants (regions ①-④ in the figure 2) with equal probability ¼. When $Z$ falls in the first quadrant (region ①, probability ¼), $LRT = \|Z\|^2$ (because $\hat{\theta}$ is simply $Z$ itself, so that $\left\|Z - \hat{\theta}\right\|^2 = 0$), which is distributed as $\chi_2^2$; and both $\hat{\theta}_1$ and $\hat{\theta}_2$ are

positive. When Z lies in the second quadrant, $\left(\hat{\theta}_1, \hat{\theta}_2\right) = \left(0, Z_2\right)$, so that $LRT = \|Z\|^2 - Z_1^2 = Z_2^2 \sim \chi_1^2$, and only $\theta_2$ is estimated positive. Symmetric logic applies to the fourth quadrant, so that $LRT = Z_1^2 \sim \chi_1^2$, and only $\theta_1$ is estimated positive. When Z lies in the third quadrant, the projection $\left(\hat{\theta}_1, \hat{\theta}_2\right) = \left(0, 0\right)$, so that $LRT = \|Z\|^2 - \|Z\|^2 = 0$ and both parameters are estimated zero. In summary, LRT is distributed as the mixture $\frac{1}{4}\chi_0^2 + \frac{1}{2}\chi_1^2 + \frac{1}{4}\chi_2^2$, with the different mixture components determined by the number of parameters estimated positive.

Figure 2. The parameter space $\theta \in \left\{\left(\theta_1, \theta_2\right) : \theta_1 > 0, \theta_2 > 0\right\}$

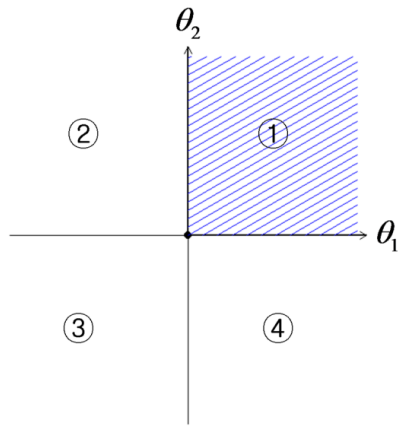

The above derivation can be summarized in other words as follows. For $d = 0, 1, 2$, conditional on the number of variance parameters estimated positive being $d$, say, the likelihood ratio test statistic should have the asymptotic distribution $\chi_d^2$. The signs of $Z_1$ and $Z_2$ are equally likely to be any of the four combinations (+,+), (+,-), (-,+), or (-,-), so the number of variance parameters estimated positive follows the binomial distribution $B(2, \frac{1}{2})$. So the asymptotic distribution should be a mixture of chi-square distributions, with the mixing probability for $\chi_d^2$ being the binomial probability $\binom{2}{d} 2^{-d}$.

The above reasoning, however, ignores the covariance parameter $a_{12}$. For simplicity, let us focus our attention on the parameters of interest $a_{11}, a_{22}$, and $a_{12}$ and ignore the nuisance parameters in the model that describe the polygenic and environmental effects. The single factor model is constrained according to $a_{12} = a_1 a_2$. That is, letting $\theta = \left(\theta_1, \theta_2, \theta_3\right) = \left(a_{11}, a_{22}, a_{12}\right) = \left(a_1^2, a_2^2, a_1 a_2\right)$, the constraint in the single factor model is expressed by $\theta_3 = \pm\sqrt{\theta_1 \theta_2}$. So for this problem, the null hypothesis is $\Omega_0 = \{\theta \in \mathbb{R}^3 : \left(\theta_1, \theta_2, \theta_3\right) = (0, 0, 0)\}$ and the alternative hypothesis is $\Omega_1 = \{\theta \in \mathbb{R}^3 : \theta_1 \geq 0, \theta_2 \geq 0, \theta_3 = \pm\sqrt{\theta_1 \theta_2}\}$. Here we are thinking of the parameter space as a curved surface in three dimensional space, not just a single quadrant in the two dimensional space $\{\left(a_{11}, a_{22}\right)\}$. Figures 3(a) and 3(b) shows this surface from different angles. Under the null hypothesis, $\theta_1$ and $\theta_2$ are on the boundary of their respective ranges (the nonnegative numbers), while $\theta_3$ is not on the boundary of its range, since it can take both positive and negative values. To apply Theorem 3 of Self and Liang (1987) to this problem, we note that the hypotheses $\Omega_0 = \{\theta = (0,0,0)\}$ and $\Omega_1$ are already cones, so that the approximating cones are $C_{\Omega_0} = \Omega_0$ and $C_{\Omega_1} = \Omega_1$. Suppose for illustration that the information matrix under the null hypothesis, $I(0,0,0)$, is the identity matrix $I = I_3$. A general case will be discussed in Section 4.3.

The above problem arises from the original genetic problem by using its asymptotic equivalence to a problem with a single observation from a Gaussian distribution, ignoring nuisance parameters, and considering the special case where the information matrix is the identity. For future reference let us refer to this simplified version as the 3-dimensional asymptotic problem.

Assume the null hypothesis holds. As in (4), the asymptotic distribution of the LRT is the same as the distribution of $LRT = \inf_{\theta \in \Omega_0} \|Z - \theta\|^2 - \inf_{\theta \in \Omega_1} \|Z - \theta\|^2 = \|Z\|^2 - \inf_{\theta \in \Omega_1} \|Z - \theta\|^2 = \|Z\|^2 - \|Z - \hat{\theta}\|^2$, where $Z \sim N(0, I)$ and $\hat{\theta} = (\hat{\theta}_1, \hat{\theta}_2, \hat{\theta}_3)$ is the vector $\theta$ that achieves the infimum in $\inf_{\theta \in \Omega_1} \|Z - \theta\|^2$, that is, the projection of $Z$ onto the alternative hypothesis surface $\Omega_1$.

The distribution of LRT is a mixture of components that are distinguished by the number of variance parameters $\hat{\theta}_1$ and $\hat{\theta}_2$ that are estimated positive (which, by the constraint that defines $\Omega_1$, also determines whether or not $\hat{\theta}_3$ is estimated nonzero). As before, let $\nu$ be the number of variance parameters $\hat{\theta}_1$ and $\hat{\theta}_2$ that are estimated positive, that is, the number of positive components among $\hat{\theta}_1$ and $\hat{\theta}_2$; the possible values for $\nu$ are 0, 1, and 2.

The determination of $\nu$ can be explained using the polar cone of the parameter space $\Omega_1$. The polar cone of $\Omega_1$ is defined to be the set $P_{\Omega_1} = \left\{ Z \in \mathbb{R}^3 : Z \cdot \theta \leq 0 \ \ \forall \theta \in \Omega_1 \right\}$. Figures 3(c) depicts the boundary of the polar cone of the parameter space.

When $Z$ is in the polar cone, we claim $\nu = 0$. To see this, assume $Z \in P_{\Omega_1}$ and note that if $\theta \in \Omega_1$ then $Z \cdot \theta \leq 0$ by definition of the polar cone, which implies $\|Z - 0\|^2 \leq \|Z - \theta\|^2$. Thus, the origin is the point in $\Omega_1$ that is closest to $Z$, which gives $\nu = 0$.

When $Z$ lies outside the polar cone, we claim $\nu = 2$ with probability one. First, by definition, for any point $Z$ outside the polar cone $P_{\Omega_1}$, there exists a vector in $\Omega_1$ that has positive inner product with $Z$. This implies the projection of $Z$ onto that vector is closer to $Z$ than the origin, proving $\nu \neq 0$. Second, we want to show $\nu \neq 1$ with probability one. Note that $\nu = 1$ can occur when $Z$ lies on the $Z_3 = 0$ plane in the first, second, and fourth quadrants, which has probability zero. We claim that whenever $Z_3$ is nonzero (which occurs with probability 1), $\nu$ cannot be 1. This is shown in each quadrant in the following paragraphs.

Figure 3. Plots (a) and (b) are the parameter space of points $\theta = (\theta_1, \theta_2, \theta_3)$ such that $\theta_3 = \pm\sqrt{\theta_1\theta_2}$. Plot (c) shows the boundary of the polar cone of the parameter space in blue. Plot (d) shows the points $Z$ that give small LRT values ($\leq 0.02$) in green. Plot (e) shows the points $Z$ giving $\nu = 0$ when $Z \sim N(0, I)$. Plot (f) shows the points $Z$ with $\nu = 0$ when the covariance of $Z$ is not the identity matrix, but a general covariance matrix obtained from the Fisher information matrix of a genetic model

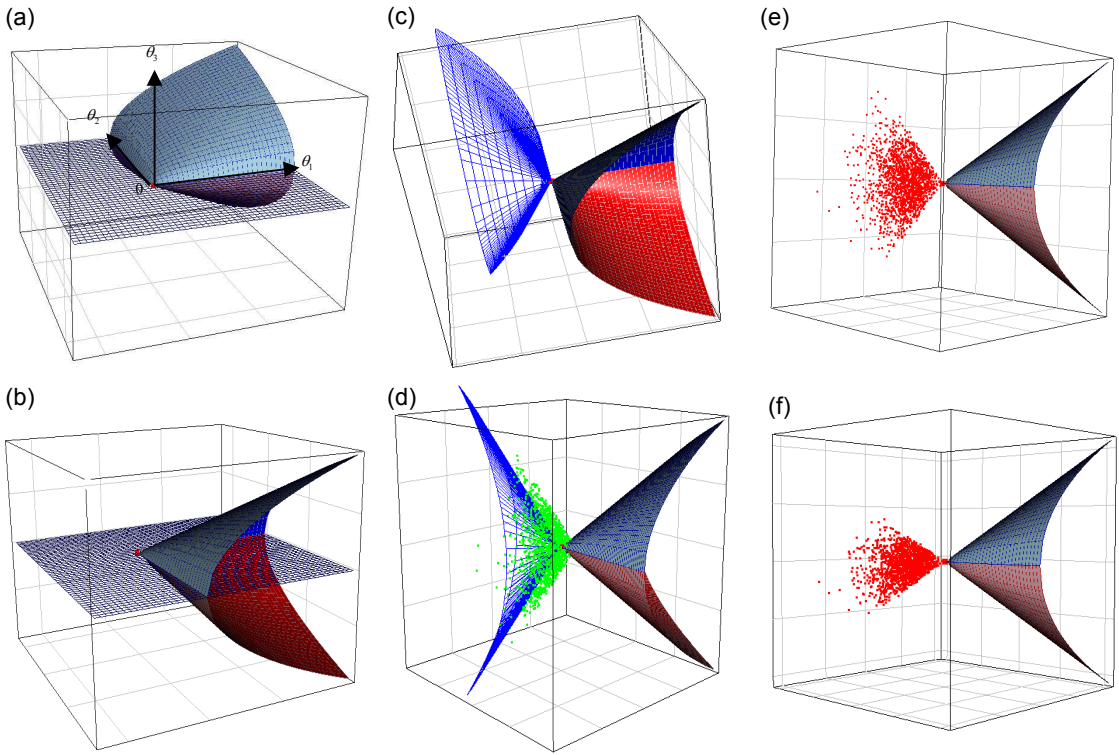

Suppose $(Z_1, Z_2)$ falls in the first quadrant, that is, $Z_1 > 0$ and $Z_2 > 0$. Without loss of generality, suppose $Z_3 > 0$ (an analogous argument will handle the case $Z_3 < 0$) and $Z_1 > Z_2$. Then the point in $\Omega_1$ having $\nu = 1$ that is closest to $Z$ is $(Z_1, 0, 0)$. However, it is obvious that for sufficiently small $\varepsilon > 0$, the point $(Z_1, \varepsilon, \sqrt{Z_1\varepsilon}) \in \Omega_1$, which has $\nu = 2$, is closer to $Z$ than $(Z_1, 0, 0)$ is, proving that $\nu$ cannot be 1. Thus we have shown that when $(Z_1, Z_2)$ falls in the first quadrant, we have $\nu = 2$ with probability 1.

Suppose $(Z_1, Z_2)$ falls in the second quadrant, that is, $Z_1 < 0$ and $Z_2 > 0$. Again suppose $Z_3 > 0$, and observe that the point in $\Omega_1$ having $\nu \leq 1$ that is closest to $Z$ is $(0, Z_2, 0)$. Define $\theta(\varepsilon) = (\varepsilon, Z_2, \sqrt{\varepsilon Z_2})$, and define $D(\varepsilon)$ to be the squared distance $\|Z - \theta(\varepsilon)\|^2 = (Z_1 - \varepsilon)^2 + (Z_3 - \sqrt{\varepsilon Z_2})^2$. The derivative $D'(\varepsilon) = 2(\varepsilon - Z_1) + Z_2 - Z_3\sqrt{Z_2 / \varepsilon}$ approaches $-\infty$ as $\varepsilon \downarrow 0$. Therefore, for sufficiently small positive $\varepsilon$, we must have $D(\varepsilon) < D(0)$, so that the point $\theta(\varepsilon)$ is closer to $Z$ than $\theta(0) = (0, Z_2, 0)$ is, which implies that $\nu \leq 1$ cannot hold. The case where $(Z_1, Z_2)$ lies in the fourth quadrant is analogous to the case of the second quadrant. Thus, for both the cases of the second and fourth quadrants, we have $\nu = 2$ with probability 1.

In summary, when $Z$ lies inside the polar cone, $\nu = 0$, and when $Z$ lies outside the polar cone, $\nu = 2$ with probability one. To illustrate this, we generated 10,000 vectors $Z$ from the 3-dimensional Gaussian distribution with mean zero and identity covariance matrix. For each $Z$, we calculated $\nu$ by estimating $\hat{\theta}$ with the optimization being performed under the constraints $\theta_1 \geq 0, \theta_2 \geq 0, \theta_3 = \pm\sqrt{\theta_1\theta_2}$ using the program Mx, which applies here because the projection $\hat{\theta}$ is also the MLE in this problem. The red points in Figure 3(e) are the points $Z$ having $\nu = 0$, which form the polar cone of the parameter space.

## 4.2.    Departures from chi-squared distributions

In Section 3.2, one of the departures from the anticipated $\chi^2$ distributions we observed was an excessive number of LRT values near 0. These small LRT values result from points $Z$ that are located close to the boundary of the polar cone, where the difference between the squared distance from $Z$ to the origin and the squared distance from $Z$ to the parameter space is small. To visualize this, using the same method of generating $Z$ as described in the previous subsection, we simulated 10,000 vectors $Z$ from $N\left(0, I_3\right)$, and for each $Z$ we performed the LRT for testing $H_0 : \theta = 0$ *vs.* $H_1 : \theta \neq 0$ with the optimization being performed under the constraints $\theta_1 \geq 0, \theta_2 \geq 0, \theta_3 = \pm\sqrt{\theta_1\theta_2}$. Then we identified the points $Z$ showing small LRT values ($\leq 0.02$), and plotted them as green in Figure 3(d). We see they are located very close to the outer surface of the polar cone.

Another type of departure we observed from the simulation results, particularly in models with larger numbers of traits, was a pattern that the distribution was shifted to the right compared to the anticipated chi-square distribution. This can be explained by the fact that the parameter space consists of two surfaces joined at an angle of less than $180°$. The angle between the two surfaces (one blue and the other red; see Figure 3 plot (a) and plot (b)) in our parameter space $\theta_3 = \pm\sqrt{\theta_1\theta_2}$ is $70.54°$ in the direction $\theta_1 = \theta_2$. When $Z$ lies between these two surfaces, the projection of $Z$ onto the parameter space involves a choice of whether the point is closer to one surface or the other, and the availability of this choice affects the distribution of LRT. An analogous example for illustration in two dimensions can help explain this problem simply. Figure 4 shows two different cases. In the first, the parameter space is a one dimensional straight line, and in the second, the parameter space is two lines joined at the origin with an angle less than $180°$, say, $90°$. For both hypothesis testing problems, the null hypothesis is the origin. For the first case, the LRT is exactly $Z_2^2 \left(= \left(Z_1^2 + Z_2^2\right) - Z_1^2\right)$ which is distributed as $\chi_1^2$. On the other hand, in the second plot, consider the points $A$ and $B$ lying between the two lines forming the parameter space. The point $A$ has LRT $= Z_2^2$ while the point $B$ has LRT $= Z_1^2$; the choice of whether LRT is $Z_1^2$ or $Z_2^2$ depends on whether the point is closer to the first line or the second. Thus, in the first quadrant, LRT is max($Z_1^2, Z_2^2$), which in general is larger than, say, $Z_1^2$, which is distributed as $\chi_1^2$. Thus, in the first quadrant, LRT does not have a $\chi_1^2$ distribution, but rather a distribution that is stochastically larger than $\chi_1^2$. Note if we changed the parameter space by decreasing the angle between the lines, the LRT in the region between the lines becomes closer to $Z_1^2 + Z_2^2$, which is distributed as $\chi_2^2$. Returning to the three dimensional problem, as the angle between the surfaces becomes closer to $0°$, the distribution of LRT in the region between the surfaces would be closer to $\chi_3^2$ because the difference between the squared distance from $Z$ to the origin, which is $Z_1^2 + Z_2^2 + Z_3^2$, and the squared distance from the $Z$ to the alternative space, which is close to zero, would be close to $Z_1^2 + Z_2^2 + Z_3^2$, which is distributed as $\chi_3^2$.

Figure 4. Influence on LRT of the angle joining the parts of the parameter. The bold lines form the parameter space and the origin is the null hypothesis.

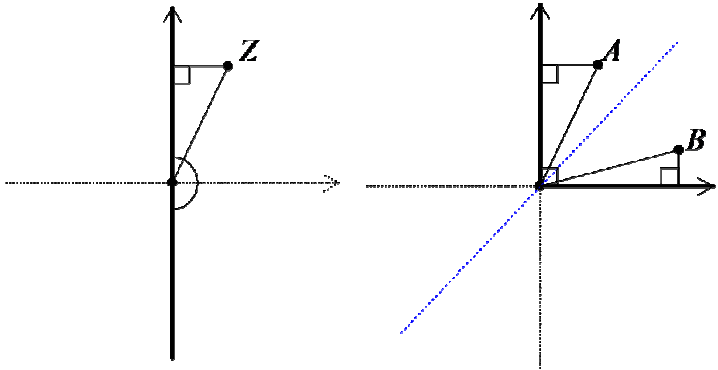

### 4.3.    General case of $I(\theta_0) \neq I$ and genetic models

#### 4.3.1.    General case of $I(\theta_0) \neq I$

So far, for simplicity, we have focused on the special case where the Fisher information $I(\theta_0)$ was the identity matrix $I$. For the general case when $I(\theta_0) \neq I$, the results of the Self and Liang theorem stated in (3) show that the asymptotic distribution of the LRT can be analyzed as above (involving a difference between the squared distance from $Z$ to the origin and the squared distance from $Z$ to the parameter space), with the Euclidean distance replaced by the more general Mahalanobis distance $\|Z - \theta\|_{\mathcal{M}}^2 = (Z - \theta)^T I(\theta_0)(Z - \theta)$. This distance takes into account the covariance of $Z$, where $Z \sim N\left(0, \, I^{-1}(\theta_0)\right)$. For the multivariate normal model, $I(\theta_0)$ can be obtained from the formula (Searle *and others*, 1992)

$$I_{ij}(\theta) = -E\left(\frac{\partial^2 l(\theta)}{\partial \theta_i \partial \theta_j}\right) = \frac{1}{2} Tr\left(\Sigma^{-1} \frac{\partial \Sigma}{\partial \theta_i} \Sigma^{-1} \frac{\partial \Sigma}{\partial \theta_j}\right). \tag{5}$$

#### 4.3.2.    Generating the asymptotic null distribution in a genetic model

The results in (3) and (5) above allow us to simulate from the asymptotic null distribution in genetic models that have general Fisher information matrices. For example, we did this for the $k = 2$ trait genetic model specified in Section 3, that is, the model with parameter values $g_{11} = g_{22} = 0.4$, $e_{11} = e_{22} = 0.6$, $e_{12} = 0.04$ and $e_{12} = 0.06$. After obtaining $\Sigma$ by plugging in the specified values in $G$ and $E$, using (5) we calculated the $9 \times 9$ Fisher information matrix for the full set of parameters $\left(a_{11}, \, a_{22}, \, a_{12}; \, g_{11}, \, g_{22}, \, g_{12}; e_{11}, \, g_{22}, \, g_{12}\right)$. After extracting the $3 \times 3$ submatrix $V$ of the inverted Fisher information matrix corresponding to the parameters of interest $\theta = \left(a_{11}, \, a_{22}, \, a_{12}\right)$, we generated 10,000 vectors $Z$ from $N(0, V)$. For each $Z$, we performed the LRT for testing $H_0 : \theta = 0$ versus $H_1 : \theta \neq 0$ under the constraints $\theta_1 \geq 0$, $\theta_2 \geq 0$, $\theta_3 = \pm\sqrt{\theta_1\theta_2}$ using Mx. In this application of Mx, we assigned fixed values for the known covariances in $V$, and Mx optimized the likelihood only over mean parameters. We also determined $\nu$ using the same technique of comparing a full model to a partially constrained model as described in Section 3.

According to the results, the mixing proportions for $\nu = 0$, $\nu = 1$, and $\nu = 2$ were 0.1477, 0, and 0.8523, and critical values for $\alpha = 0.01$ and $\alpha = 0.05$ were 8.713 and 5.405, which were quite similar to the results in Table 2. To confirm the results in this section were consistent with the genetic simulation results for the $k = 2$ trait null distribution in Section 3, a Kolmogorov-Smirnov test gave a P value of 0.2063.

Plot (f) in Figure 3 shows the points $Z$ with $\nu = 0$ colored as red, filling the polar cone. The contrast between the polar cones in plots (e) and (f) illustrates the effect of having a Fisher information matrix $I(\theta_0) \neq I$.

### 4.3.3. Different sets of nuisance parameter values

It is of interest to know whether and how the asymptotic null distribution is influenced by different choices of the nuisance parameters. One would expect that the asymptotic distribution could be influenced in this nonstandard problem, because as shown in (5), the Fisher information is a function of the parameter values through $\Sigma$, and the asymptotic mixing proportions discussed above are clearly influenced by the Fisher information.

To investigate this, we selected various sets of parameter values for each $k$-trait model. Since we were interested in the null distributions, the major gene covariance matrix $A$ was fixed at 0, and we needed choices for the remaining nuisance parameters in $G$ and $E$. We first chose two different polygenic models with (1) 20% effect size, i.e. $g_{11} = \cdots = g_{kk} = 0.2$, and (2) 40% effect size, i.e. $g_{11} = \cdots = g_{kk} = 0.4$ (thus environmental effects sizes are 80% and 60% respectively), to use with each multivariate model, for $k$ =2,3,4,5. Then for each given polygenic effect size, we chose five different sets of off-diagonal covariance parameters for both $G$ and $E$. In order to choose matrices that were well spread out over the range of possible covariance matrices, we generated many random covariance matrices and then used the centers of clusters found by K-means clustering as our representative matrices. We used the inverse Wishart distribution $W^{-1}(V, d)$, where $V$ is a scale matrix and $d$ is a degrees of freedom parameter, for generating covariance matrices, with the particular choices $V = I_k$ (the $k \times k$ identity matrix) and $d = k + 1$. These choices gave uniformly distributed correlation parameters for both $G$ and $E$ (Gelman and Hill, 2006). For each simulation run in a given $k$ - trait multivariate model, we sampled two matrices $G$ and $E$ independently from $W^{-1}(I_k, k+1)$, then scaled each matrix so it had the exact pre-assigned polygenic and environmental effect sizes to extract the scaled covariance parameters. We repeated this procedure 10,000 times to obtain 10,000 sets of uniformly sampled covariance parameters. Then we applied K-means clustering with five centers to get five sets of covariance parameters, each set of parameters being the center of its cluster. For each set of nuisance parameter values obtained this way, we conducted the same method described in Section 4.3.2 to generate the asymptotic null distribution. In order to assess whether the asymptotic null distributions generated using different sets of nuisance parameters were different, we conducted pairwise Kolmogorov-Smirnov tests.

For the 10 different models for $k = 2$, the smallest of the 45 pairwise Kolmogorov-Smirnov test P values was 0.111, indicating none was different from the rest. The results for $k = 3$ traits showed similar results, where the smallest of 45 Kolmogorov-Smirnov test P values was 0.2106.

For higher numbers of traits, however, a few models were found to have large departures from the rest of the models. For $k = 4$ traits, for example, one model showed P values as small as $2.20 \times 10^{-16}$ when tested against the other models. Figure 5 shows the distributions for two of these models that tested significantly different; we see that one distribution is shifted to the right compared to the other, although not by a very large amount.

Figure 5. Simulations of asymptotic null distributions of tests under two different models with $k = 4$ traits. Each plot shows the $\nu = 4$ component of the corresponding asymptotic distribution. The blue curves show the $\chi_4^2$ density, included only to aid visual comparison of the two red distributions.

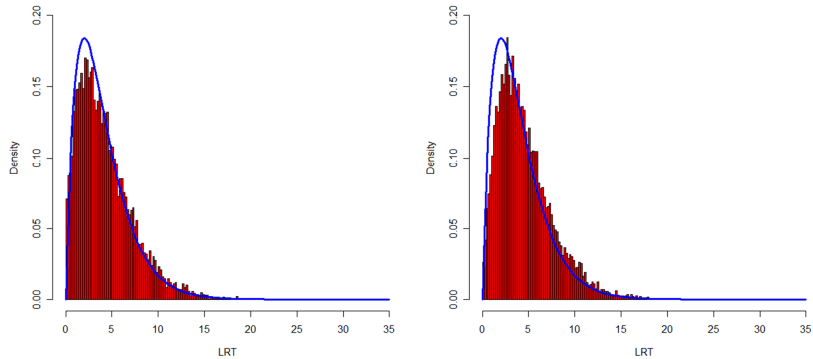

To see how actual P values would be influenced by the distributions of different models, we estimated P values corresponding to the six LRT values 1, 3, 10, 15, 18, and 20 for each of ten models for $k = 4$. According to the results in Table 3, models 5, 6 and 8 (which were the ones that previously showed significant departures from the rest of the models) had larger P values for the smaller ($\leq 10$) LRT values, and P values that were not distinctly different from those of the other models for larger LRT values ($\geq 15$).

Table 3. P values from 10 different asymptotic null distributions of 10 genetic models for $k = 4$ tests. Models that showed significant P values for pair-wise Kolmogorov-Smirnov tests were marked with *.

| LRT value | Models | | | | | | | | | |
|---|---|---|---|---|---|---|---|---|---|---|
| | 1 | 2 | 3 | 4 | 5* | 6* | 7 | 8* | 9 | 10 |
| 1 | 0.9013 | 0.9000 | 0.8963 | 0.9056 | 0.9409 | 0.9184 | 0.9057 | 0.9158 | 0.8969 | 0.903 |
| 3 | 0.5867 | 0.5918 | 0.5895 | 0.5960 | 0.6492 | 0.6229 | 0.5912 | 0.6209 | 0.5923 | 0.5945 |
| 10 | 0.0505 | 0.045 | 0.0476 | 0.0548 | 0.0591 | 0.0544 | 0.0514 | 0.0527 | 0.0521 | 0.0511 |
| 15 | 0.0051 | 0.0045 | 0.0058 | 0.0072 | 0.0067 | 0.0059 | 0.0066 | 0.0055 | 0.007 | 0.0056 |
| 18 | 0.0018 | 0.0008 | 0.0021 | 0.0018 | 0.0012 | 0.0014 | 0.0013 | 0.0015 | 0.002 | 0.0019 |
| 20 | 0.0008 | 0.0005 | 0.0009 | 0.0009 | 0.0005 | 0.0004 | 0.0006 | 0.0006 | 0.0007 | 0.0004 |

# 5. Proposed method for generating asymptotic null distributions and calculating P values

As we have shown, asymptotic null distributions of multivariate linkage tests have several complex features: the mixing probabilities are not binomial distributions and the mixture components show severe departures from chi-square distributions. The asymptotic null distributions also varied depending on nuisance parameter values to some extent, although the effects on P values were not dramatic for larger LRT values. These complexities in the asymptotic null distribution raise challenges for the assessment of significance of multivariate linkage findings. In practice, researchers typically obtain P values for multivariate linkage tests using empirical methods such as gene-dropping, permutation, and bootstrap methods. These methods, however, involve generating large numbers of replicated data sets with large sample sizes, which can be impractical for assessing small P values and require long computation times for high precision.

Here we propose a method to calculate P values by generating from an asymptotic null distribution. The new method is much faster than other empirical methods and it gives a correct asymptotic distribution. The idea was implicitly used in Sections 4.3.2 and 4.3.3, although there we assumed knowledge of the nuisance parameter values. Here we specify a method that is applicable to the more realistic situation where we are given data and do not know nuisance parameter values.

The description of the method uses the following notation. In a $k$-trait model, the total parameter set consists of the parameters in $A = \left( a_{ij} \right)$, $G = \left( g_{ij} \right)$ and $E = \left( e_{ij} \right)$, where each matrix has $m = \dfrac{k(k+1)}{2}$ parameters. The full parameter set contains all $3m$ parameters in $A$, $G$, and $E$. The parameters in $A$ consist of the parameters of interest $\theta = (\theta_1, \theta_2, ..., \theta_m)$, with constraints of the form $\theta_i \geq 0$ for $i \leq k$ and $\theta_i = \pm \sqrt{\theta_{p(i)} \theta_{q(i)}}$ for $i > k$, where we write $\theta_i = a_{p(i), q(i)}$ for $i > k$. For example, for the models with $k = 3$ traits, the parameters of interest are $\theta = (\theta_1, \theta_2, \theta_3, \theta_4, \theta_5, \theta_6) = (a_{11}, a_{22}, a_{33}, a_{12}, a_{13}, a_{23}) = \left( a_1^2, a_2^2, a_3^2; \ a_1 a_2, \ a_1 a_3, \ a_2 a_3 \right)$, and the constraints are $\theta_1 \geq 0$, $\theta_2 \geq 0$, $\theta_3 \geq 0$, $\theta_4 = \pm \sqrt{\theta_1 \theta_2}$, $\theta_5 = \pm \sqrt{\theta_1 \theta_3}$, and $\theta_6 = \pm \sqrt{\theta_2 \theta_3}$.

## 5.1. Method

Suppose we have a given data set and have calculated a LRT statistic for a multivariate linkage test for that data set, and we wish to calculate a P value.

STEP 1: Estimate the Fisher information from the given data.

STEP 1(a): Under the null hypothesis, estimate all parameters in the model--$\hat{G}$ and $\hat{E}$. Use these to form the estimate $\hat{\Sigma} = \begin{bmatrix} \hat{G} + \hat{E} & \frac{1}{2}\hat{G} \\ \frac{1}{2}\hat{G} & \hat{G} + \hat{E} \end{bmatrix}$.

STEP 1(b): Using $\hat{\Sigma}$ and the formula (5), calculate the $(3m) \times (3m)$ Fisher information matrix for the full parameter set and invert it.

STEP 1(c): Extract the $m \times m$ submatrix of the inverted Fisher information matrix corresponding to $\theta$, the parameters of interest, and denote it as $V$.

STEP 2: Generate $N$ random vectors $Z$ from the multivariate Normal distribution $N(0, V)$. An appropriate choice of $N$ depends on the desired precision level.

STEP 3: For each $Z$, perform the LRT for the null hypothesis $H_0 : \theta = 0$ in the model $Z \sim N(\theta, V)$, under the constraints $\theta_i \geq 0$ for $1 \leq i \leq k$ and $\theta_i = \pm\sqrt{\theta_{p(i)}\theta_{q(i)}}$ for $i > k$. Here the covariance matrix $V$ is fixed at the value calculated in STEP 1(c), and the test is performed by optimizing the likelihood only over the mean parameters in the vector $\theta$.

STEP 4: The P value is calculated as the fraction of LRT values that are larger than or equal to the LRT value calculated from the given data.

An example Mx script that performs the LRT in STEP 3 above is available in the web supplement at http://www.stat.yale.edu/~sh437/mxExample.

## 5.2. Applied example in an illustrative data set

In this section, we apply our proposed method and other well known empirical methods for assessing P values to an illustrative data set and compare the times for completing the tasks. To create the illustrative data set, we simulated 2,000 independent sib-pairs with two traits generated from the bivariate model with the same parameter values used in an earlier simulation: $g_{11} = g_{22} = 0.4$, $e_{11} = e_{22} = 0.6$, $g_{12} = 0.04$ and $e_{12} = 0.06$. For each sibling, we simulated a segment of chromosome with length 80 cM, having 16 markers located 5 cM apart. Each marker had 8 equally frequent alleles and IBD sharing proportions $\pi_{i,12}$ were estimated by Merlin (Abecasis et al., 2001) with grid size 2.5 cM. We prespecified a single test point at 22.5 cM on the chromosome where the LRT would be conducted.

We applied three well known methods for obtaining empirical critical values and P-values: gene dropping (Terwilliger *and others*, 1993), permutation (Wan *and others*, 1997), and the bootstrap (Marlow *and others*, 2003). Keeping the phenotype data fixed, the gene dropping method simulates genotypes on founders' chromosomes using estimated marker allele frequencies and then segregates the chromosomes to offspring using marker recombination fraction information. In this way, there is no association between the phenotype and the markers and the generated data can be used for obtaining the null distribution of a linkage test statistic. Permutation methods fix genotypes and shuffle the trait values among individuals (or vice versa) to break the association between the phenotypes and the genotypes, still keeping the same trait distribution. A number of methods have been used, differing in the details of how genotypes or traits are permuted. For our sib-pair data set, we fixed the trait values and permuted IBD estimates among families. The (parametric) bootstrap, which is not as commonly applied in linkage analysis as the previous two methods, estimates parameters from a given data set, treats the estimates as if they were true parameters, and generates replicate data sets, keeping the genotypes the same as in the original data.

We applied the above three methods to our illustrative data set, generating 10,000 replicates for each method using R and running likelihood ratio tests at the pre-specified test point for each replicated data set using Mx. For STEP 2 in our proposed method, we used $N = 10,000$.

In this experiment, our method took 56 minutes to complete the task while the gene-dropping, bootstrap and permutation methods took 5893, 3182, 1611 minutes, respectively. The empirical methods took longer than our method since they require generating many replicate data sets with each data set including as many observations as in the given data set (2000 in our experiment), whereas our method requires generating only one Gaussian observation in each data set. Also, each Mx optimization took longer for the empirical methods than for our proposed method, again because our proposed method applies Mx to data sets of just one Gaussian observation.

## 6.    Discussion

Evaluating the null distribution of a test statistic is of course a prerequisite for valid P-values. The previously stated asymptotic null distribution, which gave strongly anticonservative P-values, has been used in several studies. For example, Amos *and others* (2001) used this asymptotic null distribution for $k = 2$ traits, and SOLAR (including the current version 4.1.5), the most widely used variance component linkage software, also uses this incorrect distribution in its implementation of multivariate linkage analysis.

In fact, some researchers (Amos and de Andrade, 2001; Evans *and others*, 2004; Marlow *and others*, 2003; Monaco, 2007) have mentioned the possibility that the previously stated asymptotic null distribution might be incorrect, partially due to the inconsistent bivariate simulation results shown by Amos *and others* (2001). In a later paper, Amos and de Andrade (2001) mention this problem again, raising a question: "Theoretically, this constraint results in a 1:2:1 mixture of chi-squared distributions having, respectively, 0, 1, and 2 degrees of freedom. …the pleiotropy constrained test did not appear to follow its limiting distribution." However, clear statements confirming and quantifying these discrepancies and explaining their nature and origin have not appeared previously, and some researchers (Evans *and others*, 2004; Monaco, 2007) have expressed the desirability of clarifying the asymptotic null distribution of this test. A part of the purpose of this paper is to provide such clarification. Our simulations establish that there is indeed a real problem underlying discrepancies that had been mentioned as a puzzle in these previous papers. A device that enabled our simulations to quantify the discrepancies was to separate the mixture distribution into components defined by the number of variance parameters estimated positive. Our geometric arguments help explain how and why the discrepancies occur and give insight into the true nature of the asymptotic distribution.

It is of interest to understand, for various possible derivations that would falsely lead to the previously stated asymptotic distribution, exactly where the reasoning fails. For the simple case of $k = 2$ traits, a belief that the asymptotic null distribution is simply $0.25\chi_0^2 + 0.5\chi_1^2 + 0.25\chi_2^2$ would follow from the incorrect reasoning given in Section 4.1. This incorrect reasoning views the parameter space as a quadrant in the two dimensional space of possible values for the variance parameters, and also makes the further mistake of assuming the information matrix is the identify. Even correcting this reasoning to allow a general information matrix leads only to asymptotic distributions of the form $\left(\frac{1}{2} - p\right)\chi_0^2 + \frac{1}{2}\chi_1^2 + p\chi_2^2$;
see the case 7 in Self and Liang (1987) for the derivation of this formula. However, this mixture of chi-square distributions is still evidently incorrect, as we have seen that the correct asymptotic distribution has zero mixing probability for the $\chi_1^2$ component. All of the above reasoning is invalidated by the observation that the parameter space is actually not isomorphic to the first quadrant in two dimensions; in fact, for each point $(a_{11}, a_{22})$ in the quadrant, there are two different points in the parameter space, with one point having $a_{12} = \sqrt{a_{11}a_{22}}$ and the other having $a_{12} = -\sqrt{a_{11}a_{22}}$ .

Since there are two distinct distributions for each point in the quadrant, a reasonable attempt to correct this topological problem is to view the parameter space as two quadrants. For example, as in Figure 6, we could consider the parameters to be $a_1$ and $a_2$ where $a_1$ can take any real value (including negative numbers) and $a_2$ is constrained to be nonnegative. These two parameters determine the covariance $a_{12}$ by the relationship $a_{12} = a_1 a_2$ . More precisely, to make the problem identifiable, we should remove the ray of points $\{(a_1, a_2) : a_1 < 0,\ a_2 = 0\}$ from the parameter space, since the two points $(a_1, 0)$ and $(-a_1, 0)$ correspond to the same distribution.

Figure 6. The modified half-plane parameter space

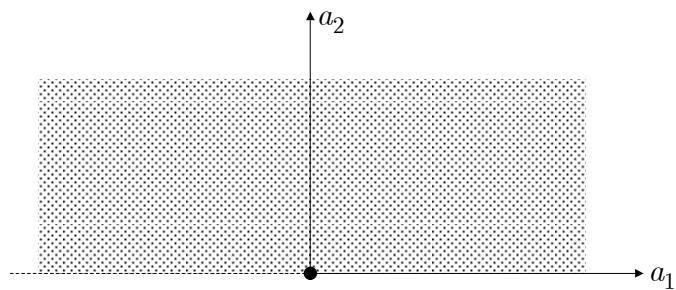

Although this modified half-plane gives a correct one-to-one correspondence with the set of distributions in our problem, attempting to apply the reasoning of Self and Liang (1987) to this situation (see their Case 6) would lead to a conjectured asymptotic null distribution of $\frac{1}{2}\chi_1^2 + \frac{1}{2}\chi_2^2$, which is also incorrect. In fact, the assumptions required for the results of Self and Liang are questionable in our problem in this two-dimensional formulation, since the modified half-plane is not closed and the information matrix at the null hypothesis $(a_1, a_2) = (0,\ 0)$ is the zero matrix.

In this paper, understanding the correct nature of the asymptotic distribution was facilitated by viewing the parameter space in a different way. Thinking of our parameters of interest as $\theta = (\theta_1, \theta_2, \theta_3) = (a_{11}, a_{22}, a_{12}) = (a_1^2, a_2^2, a_1 a_2)$ led to considering a two-dimensional surface in a three-dimensional parameter space, which led to a number of conclusions. According to our simulations, for $k = 2$ traits, there is zero mixing probability for the case where only one variance parameter is estimated positive, and highest probability for the case where both variance parameters are estimated positive. This phenomenon was displayed in our 3-dimensional asymptotic version of the problem, in which we saw that the only cases having just one variance parameters estimated positive occur when $Z$ , a single Gaussian observation, lies in either the first, second or fourth quadrant of the plane $Z_3 = 0$, which has probability zero. Also, except when $Z$ lies inside the polar cone of the parameter space, which occurs only in a subset of the third quadrant, all other locations for $Z$ lead to the case where both variance parameters are estimated positive, thus giving the highest probability to this case. We saw that a strong departure from the chi-square distribution arises from points $Z$ located close to the outer surface of the polar cone, which give small LRT values. Another type of departure from chi-square distributions was also explained by the fact that the parameter space consists of two surfaces joined at an angle of less than $180°$.

Understanding the asymptotic distribution was also helpful in developing a practical method for assessing significance in linkage findings. In contrast with a typical standard statistical problem, in which the asymptotic distribution is simple and definite (e.g. a chi-square distribution), asymptotic distributions in multivariate linkage testing are more complicated and depend on nuisance parameters. Using the asymptotic equivalence of the LRT in the genetic linkage problem and a simpler LRT based on a single Gaussian observation, our method gains speed by generating and processing only single Gaussian observations instead of full replicated data sets. The speed advantage of our method is particularly helpful for assessing highly significant linkage findings, where P values are small and the methods need to use large values of $N$ in order to estimate the P values with precision.

## 7.    Acknowledgment

We are grateful to Dr. Elena Grigorenko for inspiring our interest in these tests and for helpful advice and pointers to the literature, to Dr. Michael Neale for help with Mx, and to the Yale Center for High Performance Computation in Biology and Biomedicine (NIH grant RR19895-02). This research was partially supported by NIH grant R01 DC007665.